%% file: bg_pressure.tex
\documentclass[twocolumn,showpacs,superscriptaddress,preprintnumbers,amsmath,amssymb,aps,prl]{revtex4-1}
\usepackage{graphicx}
\usepackage{mathptmx}
\usepackage{verbatim}
\usepackage{graphicx}
\usepackage{amsmath}
\usepackage{amssymb}
\usepackage{amsfonts}
\usepackage{mathrsfs}
\usepackage{amsmath} 
\usepackage{color}
\usepackage{graphicx}
\usepackage{bm}    % bold maths\Delta_0
\usepackage{amssymb}
\usepackage{xspace}
\usepackage{hyperref}
\usepackage{float}
\usepackage{bbold}
\usepackage{tabu}
\usepackage{textcomp}
\usepackage{dcolumn}   % Align table columns on decimal point
\newcolumntype{L}[1]{>{\raggedright\arraybackslash}m{#1}}
\newcolumntype{C}[1]{>{\centering\arraybackslash}m{#1}}
\newcolumntype{R}[1]{>{\raggedleft\arraybackslash}m{#1}}
\newcolumntype{N}{@{}m{0pt}@{}}

\begin{document}
%%%%%%%%%%%%%%%%%%%%%%%%%%%%%%%%%%%%%%%%%%%%%%%%%%%

\title{Pressure Induced Compression of Flatbands in Twisted Bilayer Graphene}
\author{Bheema Lingam Chittari}
\affiliation{Department of Physics, University of Seoul, Seoul 02504, Korea}
\author{Nicolas Leconte}
\affiliation{Department of Physics, University of Seoul, Seoul 02504, Korea}
\author{Srivani Javvaji}
\affiliation{Department of Physics, University of Seoul, Seoul 02504, Korea}
\author{Jeil Jung}
\affiliation{Department of Physics, University of Seoul, Seoul 02504, Korea}
%

%%%%%%%%%%%%%%%%%%%%%%%%%%%%%%%%%%%%%%%%%%%%%%%%%%%%%%%%%%%%%%%%
\begin{abstract}
We investigate the bandwidth compression due to out of plane pressure of the moir\'e flatbands near charge neutrality in 
twisted bilayer graphene for a continuous range of small rotation angles of up to $\sim 2.5^{\circ}$. 
The flatband bandwidth minima angles are found to grow linearly with interlayer coupling $\omega$ and decrease with Fermi velocity.
Application of moderate pressure values of up to 2.5~GPa achievable through a hydraulic press should allow to access 
%the first magic angle flatband for angles as large as $\sim1.5^{\circ}$ instead of $\sim 1^{\circ}$ at zero pressure.
a flatband for angles as large as $\sim1.5^{\circ}$ instead of $\sim 1^{\circ}$ at zero pressure.
This reduction of the moir\'e pattern length for larger twist angle implies increase of the effective Coulomb interaction scale per 
moir\'e cell by about 50\% and enhance roughly by a factor of $\sim 2$ the elastic energy that resists the commensuration
strains due to the moir\'e pattern.
Our results suggest that application of pressure on twisted bilayer graphene nanodevices through a hydraulic press 
will notably facilitate the device preparation efforts required for exploring the ordered phases near magic angle flatbands.
\end{abstract}
%    73.22.Pr        Electronic structure of graphene.
%    71.20.Gj        Electronic structure of semimetal.
%    Tight-binding methods (atomic physics), 31.15.aq
%% 71.15.Mb      Density-functional theory
%% 71.15.Nc 	   Total energy and cohesive energy calculations 
%% 71.70.Gm 	   Exchange interactions 
%% 73.22.Gk      Broken symmetry phases
%\pacs{71.20.Tx, 73.61.Wp, 78.66.Tr}
\maketitle
% Lattice Constants are:  Graphene a - 2.46 A
% Boron Nitride a = 2.50 A 
% TMD's a = 3.15 A 

\section{Introduction}
%\vspace{-2pt}
%\vspace{-10pt}
The studies on the electronic properties of twisted bilayer graphene (tBLG) has received recently 
a renewed boost of interest thanks to groundbreaking discoveries of flatband superconductivity~\cite{pjarillo_superconductivity} and Mott gaps~\cite{pjarillo_mott,kyounghwan}
for certain magic angles close to $\sim1^{\circ}$ where a high density of states is generated near the 
charge neutral point.
The possibility of tailoring narrow flatbands in systems with such remarkable simplicity in composition as graphene
consisting of only carbon atoms makes it an attractive testbed for trying to understand its microscopic mechanisms 
of electron correlations and its coupling with lattice vibrations as fully as possible~\cite{paper1,paper2,paper3,paper4,paper5}.
Other van der Waals systems including trilayer graphene on hBN, transition metal dichalcogenide layers manifesting flatbands and topological superlattice bands 
have been proposed in recent literature~\cite{flatbandtmd2018allan,flatbandtmd2018randy,flatbandtmd2018kaxiras,guorui_mott2018,chittari2018,todadri}. %,suyongjung}.

% -- Decoupled
In twisted bilayer graphene the graphene layers' electronic structure can be considered effectively decoupled 
when the twist angles are large enough, typically above $\theta \gtrsim 10^{\circ}$ as the 
twisted multilayer graphene stacks grown on SiC substrates~\cite{deheer,lanzara}. 
The progressive increase in coupling is manifested in the gradual decrease of the
Fermi velocity as the twist angle is reduced starting from the Fermi velocity of graphene
for large twist angles~\cite{lopes,meiyin,sanjose,bistritzer2011,jung2014}.
%
%  -- Non perturbative
%
It is only in the regime of small twist angles on the order of $\sim1^{\circ}$ that the moir\'e patterns are long enough to 
reduce the Fermi velocity band dispersion to the point of flattening the bands almost completely due to non-perturbative 
coupling between the electronic states separated by short moir\'e reciprocal lattice ${\bf G}$-vectors in momentum space.
In this limit a series of {\em magic twist angles} of 
$\theta = 1.05^{\circ}, 0.5^{\circ}, 0.35^{\circ}, 0.24^{\circ}$, and $0.2^{\circ}$ have been predicted numerically~\cite{bistritzer2011}
that approximately follow the $\theta \sim (1/n)^{\circ}$ sequence for $n=1,2, \hdots$, 
where the band slopes assume zero values at the $\tilde{\Gamma}$ point of the moir\'e Brilouin zone, 
and the bandwidths achieve a series of minima.
% -- Sensitivity
%
One important hurdle for the realization of the flatbands in twisted bilayer graphene 
lies at the high sensitivity of electronic structure to twist near the magic angles because the bandwidths 
can undergo variations on the order of $\sim10$~meV for small variations in twist angle $\Delta \theta \sim$0.1$^{\circ}$
leading to drastic changes in the electronic properties of tBLG.

In this work we show that application of appropriate pressure to the system is an additional control knob that can control 
the flatband bandwidth of the system. %~\cite{kaxiraspressure}. 
Contrary to the twist angles whose values are determined once for all for each fabricated device, the vertical pressure 
is a continuously variable system parameter controllable with a hydraulic press~\cite{yankowitz,kaxiraspressure} 
that should be applicable also in twisted bilayer graphene devices. 
Our work presents a roadmap for tailoring flatbands in twisted bilayer graphene even when the twist angle control is not strictly precise. %
% --- Remove if needed ---
%The manuscript is structured as follows. %In Sec.~II 
We will begin by briefly introducing the model Hamiltonian used in our calculations,
%In Sec.~II 
then move on to discuss the bandwidth phase diagram of the system as a function of pressure and twist angle, 
%Further discussions on %In Sec.~III we discuss 
and further discuss the effects of an interlayer potential difference in the electronic structure of the flatbands
before closing the paper %in Sec.~IV 
with the summary and discussions.

%\section{Model Hamiltonian and moir\'e bands}
\section{Moir\'e pattern model Hamiltonian}
The Hamiltonian of twisted bilayer graphene at valley $K$ is based on the continuum model
of the graphene layer Hamiltonian~\cite{bistritzer2011} 
perturbed by stacking-dependent interlayer tunneling and intralayer potential variations~\cite{jung2014}
\begin{eqnarray}
h (\theta)= (\upsilon_{F} \hat{P}_{\theta}{\bf p} + {\bf A}({\bf r}) ) \cdot \sigma_{xy} + V({\bf r}) \mathbb{1} + \Delta({\bf r}) \sigma_z 
\end{eqnarray}
where $\sigma_{xy} = (\sigma_x, \sigma_y)$ and $\sigma_z$
are the graphene sublattice pseudospin Pauli matrices, the momentum is defined in the $xy$-plane ${\bf p} = (p_x, p_y)$
and $\hat{P}_{\theta}$ introduces a phase shift in the off-diagonal term in the Dirac Hamiltonian to account for the rotation of the layers
$e^{ \pm i \theta_{\bf k}} \rightarrow e^{\pm i (\theta_{\bf k} - \theta)} $ 
where $\theta_{\bf k}$ is measured with respect to the $x$-axis and $\theta$ is the rotation of the graphene layer with respect to $x$.
The graphene layers can be coupled through a stacking-dependent interlayer tunneling $T({\bf r})$~\cite{bistritzer2011}
resulting in the Hamiltonian
\begin{equation}
H_{\rm tBLG} = \begin{pmatrix} h(-\theta/2)  & T({\bf r})  \\ T^{\dag}({\bf r})  &  h(\theta/2) \end{pmatrix}.
\end{equation}
The symmetric and opposite rotations of the top and bottom graphene layers allows to preserve the moir\'e Brillouin zone (mBZ) orientation 
and therefore of the stacking dependent moir\'e patterns   
$V({\bf r})$, $\Delta({\bf r})$, ${\bf A}({\bf r})$ for intralayer potential variations, local mass term and 
virtual strains due to pseudomagnetic field vector potentials, 
in addition to the spatially varying interlayer tunneling $T(\bf {r})$ for every position ${\bf r}$. % as a function of local stacking configuration. 
Using the first harmonic approximation we have 
\begin{eqnarray}
M({\bf r}) = \sum_{m=1}^{6} e^{i {\bf G}_m {\bf r}} M_{m} 
\label{moirepattern}
\end{eqnarray}
where the moir\'e patterns $M({\bf r})$ can be modeled through its Fourier coefficients $M_m$. 
More explicitly, for a triangular lattice we can write the scalar and vector moir\'e patterns as~\cite{jung2014,laksono2017} 
\begin{eqnarray}
V({\bf r}) &=& 2 C_V \, {\rm Re}\left[ e^{i \phi_V} f({\bf r}) \right], \,\,\,  \Delta ({\bf r}) = 2 C_{\Delta} \, {\rm Re}\left[ e^{i \phi_{\Delta}} f({\bf r}) \right]\\
{\bf A}({\bf r}) &=&  2 C_{AB} \, \hat{z} \times \nabla {\rm Re}\left[ e^{i \phi_{xy}} f({\bf r}) \right]
\end{eqnarray}
where we have used the auxiliary function $f({\bf r}) = \sum_{m=1}^6 e^{i {\bf G}_m \cdot {\bf r}} (1 + (-1)^m )/2$,
where the six first shell mBZ recirpocal lattice vectors are ${\bf G}_m = \hat{R}_{2\pi(m-1)/3} {\bf G}_1$ for $m$ indices 
running from 1 to 6 are generated through successive rotation by $2\pi/3$ of the vector ${\bf G}_1 \simeq (0, 4\pi \theta / \sqrt{3} a)$, where $a=2.46~\AA$ is the lattice constant of graphene.
The moir\'e pattern Hamiltonian parameters obtained from {\em ab initio} calculations in sublattice representation for rigid bilayer graphene are~\cite{jung2014}
$C_{AA} = C_{B'B'} = 1.10 \, {\rm meV}$, $\varphi_{AA} = \varphi_{B'B'} = 82.54^{\circ}$,
$C_{BB} = C_{A'A'} = C_{AA}$, $\varphi_{BB} = \varphi_{A'A'} = -\varphi_{AA}$,
$C_{AB} = 2.235 \, {\rm meV}$, $\varphi_{AB} = 0^{\circ}$,
and with LDA out of plane relaxation we have
$C_{AA} = 2.3 \, {\rm meV}$, $\varphi_{AA} = 27.5^{\circ}$,
$C_{BB} = C_{AA}$, $\varphi_{BB} = -\varphi_{AA}$,
$C_{AB} = 2.08 \, {\rm meV}$, $\varphi_{AB} = 0^{\circ}$, where we use the notation $\phi_{xy} = \pi/6 - \varphi_{AB}$.
In our modeling of the rigid twisted bilayers the intralayer moir\'e patterns have a small effect in the electronic structure and they can be neglected,
whereas the relaxed moir\'e pattern parameters lead to particle-hole symmetry breaking as we will show later on.
The momentum conservation condition in twisted bilayer graphene 
\begin{eqnarray} 
{\bf k}' = {\bf k} + {\bf G}
\end{eqnarray} 
implies that a Bloch state with momentum ${\bf k}$ from one layer scatters to  ${\bf k}'$ at the other layer
through a moir\'e reciprocal lattice vector ${\bf G}$~\cite{jung2014}. 
In the small angle approximation we have
${\bf G} \simeq - \theta \hat{z} \times {\bf g}$ where the reciprocal lattice vector of graphene is represented through ${\bf g}$. 
If we consider the ${\bf q} = {\bf k} - {\bf K}$ and ${\bf q}' = {\bf k}' - {\bf K}'$ momenta measured respect to 
the Dirac points of each layer relatively displaced by $\Delta {\bf K} = {\bf K}' - {\bf K}= 2K \sin(\theta/2)$
we have the relationship ${\bf q}' = {\bf q} + {\bf K} - {\bf K}' + {\bf G} = {\bf q} + {\bf Q}$,
where the three ${\bf Q}_j$ vectors given by ${\bf Q}_0 = K \theta (0, -1) $ and 
${\bf Q}_{\pm} = K \theta (\pm \sqrt{3}/2, 1/2)$ in the small angle approximation to represent the interlayer coupling Hamiltonian 
\begin{eqnarray}
T({\bf r}) = \sum_{j} e^{ -i {\bf Q}_j {\bf r}} T^{j}_{s, s'},   \label{interlayercoupling}
\end{eqnarray}
%\begin{eqnarray}
%T^j =  \omega \sum_{j} e^{ -i {\bf Q}_j {\bf r}} T^{j}_{s, s'},
%\end{eqnarray}
where the interlayer coupling matrices are given by
\begin{equation}
T^0 =  \omega \begin{pmatrix} 1  & 1  \\ 1  &  1 \end{pmatrix},    \,\,   T^{\pm} =  \omega \begin{pmatrix} 1  & e^{\mp i 2\pi/3}  \\ e^{\pm i 2\pi/3}  &  1 \end{pmatrix}.
\end{equation}
We note that these $T^{j}$ matrices result when the twist is applied to a bilayer with $\tau = (0,0)$ initial AA stacking configuration
and differ by an additional phase of $e^{-i {\bf G}_j \tau}$ with respect to the AB stacking case where $\tau = (0,a/\sqrt{3})$. 
%used in Ref.~\cite{bistritzer2011}.    
%
%
%
% ----------------------------------------------------
%\subsection{Moir\'e bands and moir\'e Brilloiuin zone}
%
%
%Let us recall that when the original bands of graphene are folded into the superlattice bands 
%the interlayer coherence is often manifested as avoided crossings or secondary Dirac points near the mBZ boundaries. 
%For instance, the van Hove singularities that can intuitively be pictured as following from the overlap of the two Dirac cones 
%of each graphene layer forming a saddle point is found to evolve roughly proportionally with twist angle $\theta$ through
%$E_{\rm vHS} \sim \hbar \upsilon_F K_D \theta$ \cite{lopes,evaandrei,wong2015}.
%It is then natural to expect that all three aforementioned parameters of the Hamiltonian namely the Fermi velocity, interlayer coupling strength and twist angle will affect the flatband bandwidth phase diagram of twisted bilayer graphene.  
% -----------------------------------------------------
%
%
%
The smooth variation of $T({\bf r})$ in Eq.~(\ref{interlayercoupling}) can be traced back to the relatively larger 
interlayer distance of $c \sim 3.35~\AA$ when compared to 
inter-carbon distances of $a_{\rm CC} \sim 1.42~\AA$~\cite{bistritzer2011} and effectively implies that the bilayer graphene
interlayer coupling strength can be described by a single parameter $\omega = t_1/3 \sim 0.113$~eV when $c=3.35~\AA$,
and somewhat weaker $\omega \sim 0.1$~eV when out of plane LDA relaxations are allowed between the layers~\cite{jung2014}.
We note that $\omega$ is proportional to the interlayer coupling term $t_1 \sim 3 \omega$ of commensurate bilayer graphene 
evaluated at the Dirac point when only the three ${\bf G}$-vector contributions of interlayer coupling 
nearest to the Dirac point $K$ are considered.
The interlayer coupling can in principle be modeled more accurately by including  
the effects of the Fourier components for larger ${\bf G}$-vectors in momentum space 
that become more relevant in the presence of out of plane corrugations and in-plane strains~\cite{jung2014,jung2015}.
Given the high accuracy of the single parameter interlayer coupling model for describing the electronic properties of rigid twisted layers, 
we will focus on the role of $\omega$ in our precisely defined continuum model and defer the discussions about the moir\'e strains and 
larger momenta Fourier components for future work. 

\begin{figure}
\begin{center}
\includegraphics[width=8cm]{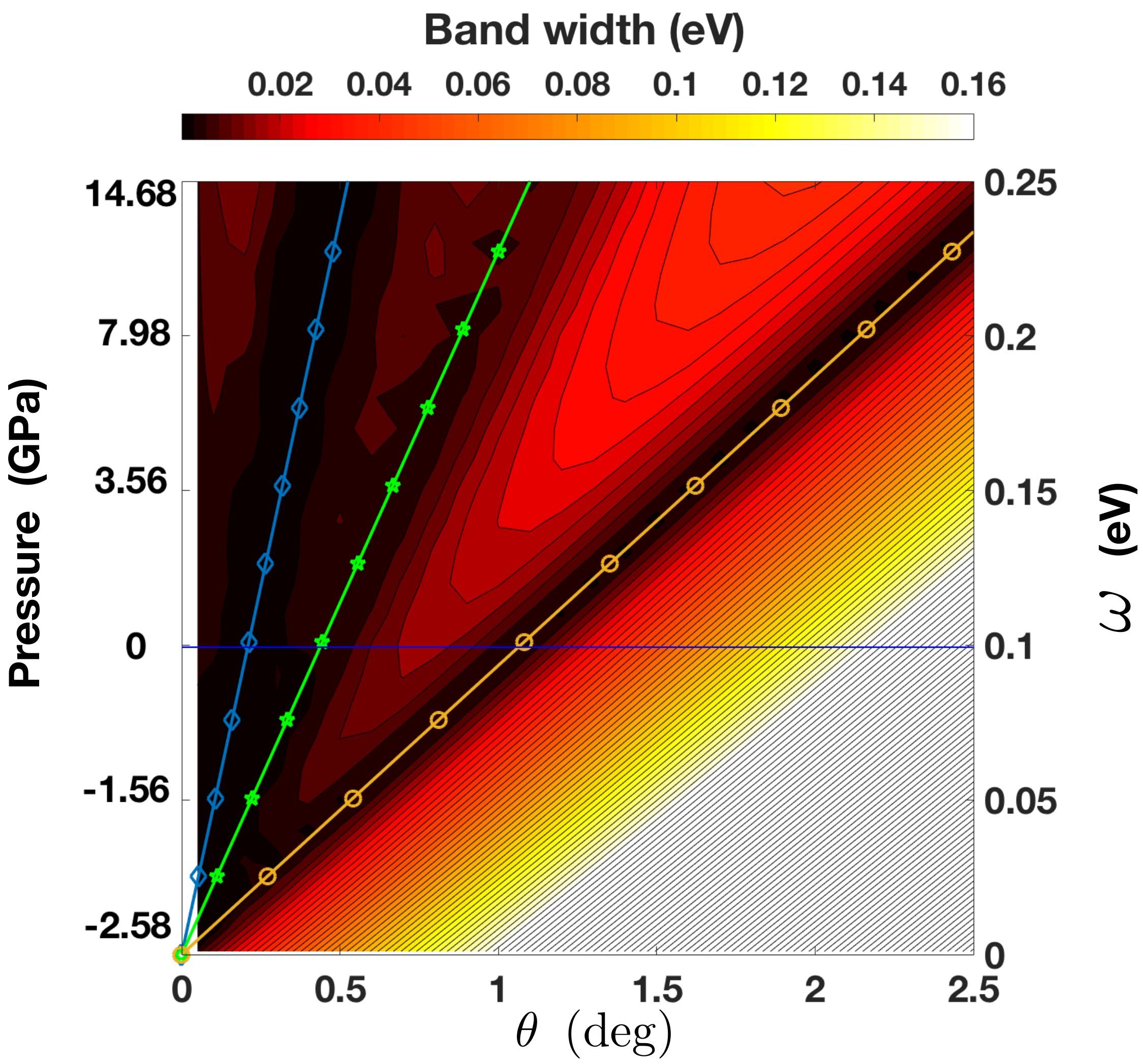}
\end{center}
\caption{
(Color online)
Colormap of the flatband bandwidth as a function of twist angle $\theta$ and interlayer tunneling strength $\omega$. 
We identify numerically straight lines in the parameter space given by 
Eq.~(\ref{mangle}), where we
represent in orange, green and blue the regions in the phase diagram corresponding to the first, second and third magic 
angles for decreasing twist angles for which bandwidth minima are achieved.
We have verified that the straight lines in $\theta$ and $\omega$ fit accurately the bandwidth minima for a variety of 
of  $\upsilon_{\rm F} =  3 |t_0 | a/2\hbar$.
%intralayer hopping terms $\left| t_0 \right|$ of equivalently Fermi velocities.
The interlayer tunneling value of $\omega \sim 0.11$~eV corresponds to rigid graphene at fixed $c=3.35~\AA$ interlayer spacing,
while out of plane relaxations within LDA reduces the effective tunneling to $\omega \sim 0.1$~eV (blue horizontal line) when no external pressure is applied. 
}
\label{figure1}
\end{figure}

\section{Pressure dependent flatband bandwidths}
In the following we obtain the phase diagram map of the bandwidth as a function of pressure
to show how increase in pressure by a few GPa can trigger the appearance of flatbands in tBLG for 
twist angles that are larger by a fraction of a degree than the magic angles at zero pressure.
We have calculated the phase diagram map of the moir\'e bands bandwidth of twisted bilayer 
graphene as a function of the three main parameters in the model Hamiltonian,
namely the Fermi velocity of each graphene layer $\upsilon_{\rm F}$, the twist angle $\theta$, 
and the pressure dependent interlayer coupling strength $\omega$. 
The Fermi velocity $\upsilon_{\rm F}$ can change with the dielectric environment due to many-body 
effects but otherwise can be assumed to be constant. 
The twist angle $\theta$ can be controlled at will during device fabrication but is difficult to modify 
afterwards in a controlled manner, whereas the pressure that controls the interlayer coupling $\omega$ can 
in principle be varied continuously in a hydraulic press up to values of 2.5~GPa~\cite{yankowitz}.
Our main results are summarized in Fig.~\ref{figure1} where we represent the flatband bandwidth colormap 
obtained as a function of interlayer coupling parameter $\omega$ and twist angle $\theta$. 
Note that the bandwidth for the low energy conduction and valence bands are electron-hole symmetric in the 
absence of intralayer moir\'e pattern terms in the Hamiltonian.
From the analysis of the numerically calculated bandwidth phase diagrams for different Fermi velocities
we find a fitting formula for the three visible flatband magic angle lines 
\begin{eqnarray}
\theta^{\circ}_{n} &=&  C_n \frac{\omega}{\left| t_0 \right|} \quad  ({\rm deg})
\label{mangle}
\end{eqnarray}
where $C_1 =  27.8$, $C_2 = 11.5$, and $C_3=5.46$,
that lead to $\theta_1 =  1.07^{\circ}$, $\theta_2 = 0.44^{\circ}$ and $\theta_3 = 0.21^{\circ}$ with $t_0=-2.6$~eV and $\omega=0.1$~eV.
These results are in fair agreement with the numerical magic angles in Ref.~\cite{bistritzer2011} 
that approximately follow the $\theta \sim (1/n)^{\circ}$ sequence 
% $\theta = 1.05^{\circ}, 0.5^{\circ}, 0.35^{\circ}, 0.24^{\circ}$, and $0.2^{\circ}$
for $t_0= - 2.7$~eV and $\omega=0.11$~eV Hamiltonian parameters,
except for the angles of $\theta = 0.35^{\circ}, 0.24^{\circ}$ which are not clearly resolved in our bandwidth phase diagram.
Our flatband magic angles are linearly proportional to the interlayer coupling $\omega$, 
and are inversely proportional to the Fermi velocity $\upsilon_{\rm F} = \sqrt{3} \left| t_0 \right| a /2\hbar$.
% or equivalently the magnitude of the intralayer hopping parameter $t_0$. 
We note that in our case the magic angles were obtained from bandwidth minima 
in the $\omega$ and $\theta$ parameter space, while previous calculations
identified the magic twist angles from the zero slope in the band dispersion at the $\tilde{\Gamma}$-point in the mBZ.
One practical implication of our findings is that it should be possible to access the flatbands by 
increasing $\omega$ with pressure by a few GPa when the twist angles in bilayer graphene are a fraction of a degree 
larger than the zero pressure magic angles.

% --- About the Fermi velocity ---
The Fermi velocity that we use as one of the free parameters in our model is an 
intrinsic property of graphene that can be enhanced by Coulomb interactions
and is therefore subject to specific environment and device quality.
Strictly speaking, a logarithmic divergence is expected for the Fermi velocity at close proximity of the Dirac point 
due to the long rangedness of the Coulomb tail which introduces $k$-dependent dispersion slope changes~\cite{guinea,chirality,jung_nonlocal}. 
Yet a constant enhanced Fermi velocity often gives an excellent fit to experimental data,
with the lower $\upsilon_F\sim 1 \times 10^6$  or $\sim 1.05 \times 10^6$ fitting well experiments of graphene 
on SiC or SiO$_2$ substrates~\cite{deheer,orlita} and CVD grown twisted bilayer graphene~\cite{dillonwong}, 
while higher $\upsilon_F\sim 1.1 \times 10^6$~m/s are better for fitting the experimental data in high quality graphene devices
with hexagonal boron nitride barrier materials~\cite{kayoung}. 
In this work, we used {\em ab initio} LDA calculation values of $t_0 = -2.6$~eV for the intralayer hopping term~\cite{ldahopping} that 
is in the lower end of the spectrum with $\upsilon_{\rm F} \sim 0.84 \times 10^6$, 
and close to  $t_0 \sim -2.7$~eV used in Ref.~\cite{bistritzer2011}.
Our tight-binding Fermi velocity choice is more appropriate for band theories that intend to introduce 
the many-body corrections explicitly on top of the non-interacting model.

\begin{figure}
\begin{center}
\includegraphics[width=8cm]{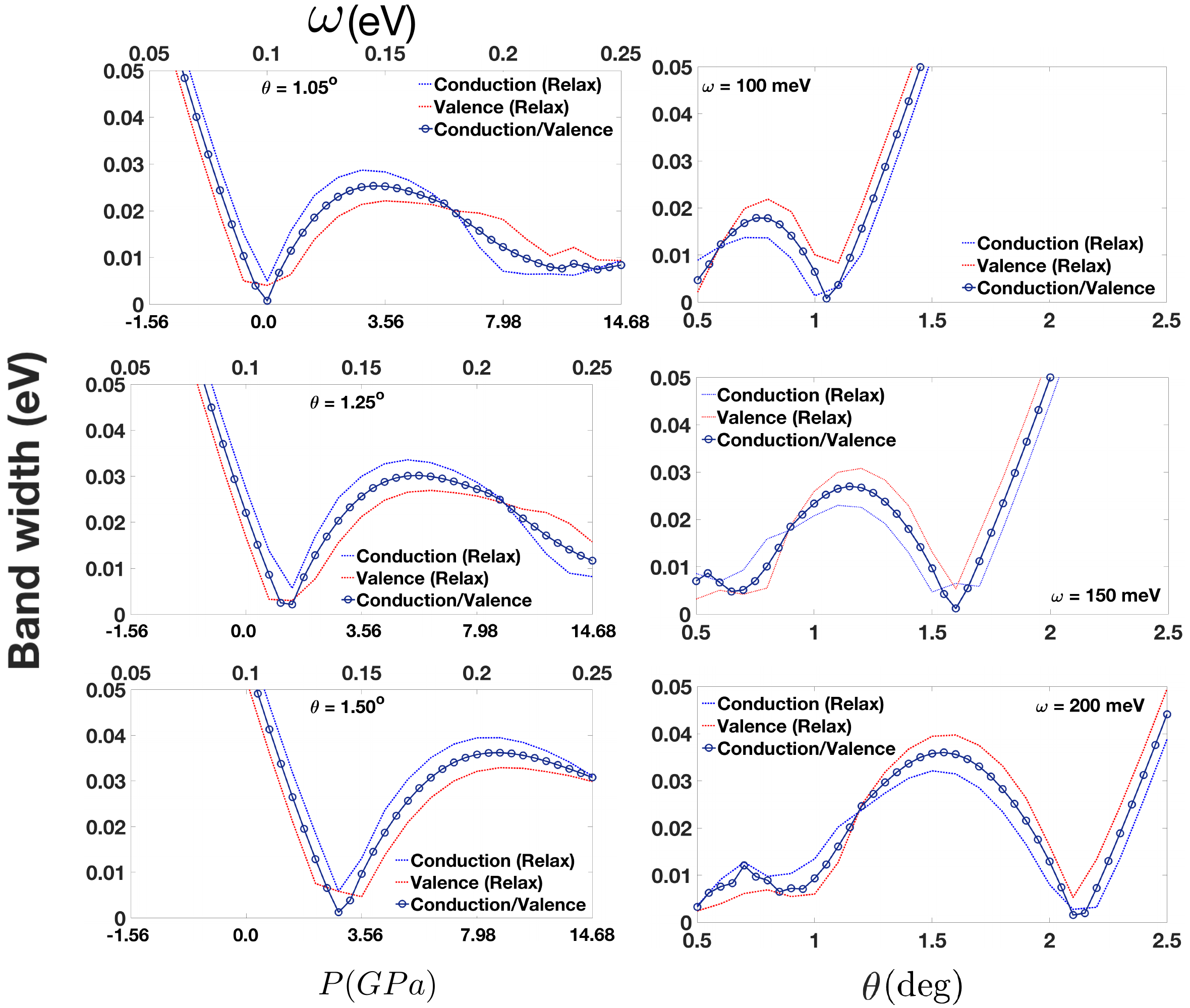}
\end{center}
 \caption{
 (Color online)  Flatband bandwidth evolution as a function of pressure and twist angle for the rigid continuum
 model with negligible intralayer moire patterns and the out of plane relaxed model that introduces a 
 small particle hole asymmetry in the low energy bands.
{\em Left Panel:} Flatband bandwidth as a function of $\omega$ (or external pressure $P$), 
for different values of the twist angle $\theta^{\circ}=1^{\circ}, 1.25^{\circ}, 1.5^{\circ}$.
We observe a progressive reduction in the bandwidth as pressure is increased from left to right
until it arrives to a minimum value. 
{\em Right Panel:} Flatband bandwidth as a function of twist angle $\theta$ for different constant
values of interlayer coupling $\omega = 100, 150, 200$~meV. 
We observe that the bandwidth has a non-monotonic dependence with respect to the twist angle
with almost vanishing bandwidth for the first magic angle but maintaining a finite value for the second magic angle.
 }
  \label{figure2}
\end{figure}

\begin{figure*}
\begin{center}
\includegraphics[width=16cm]{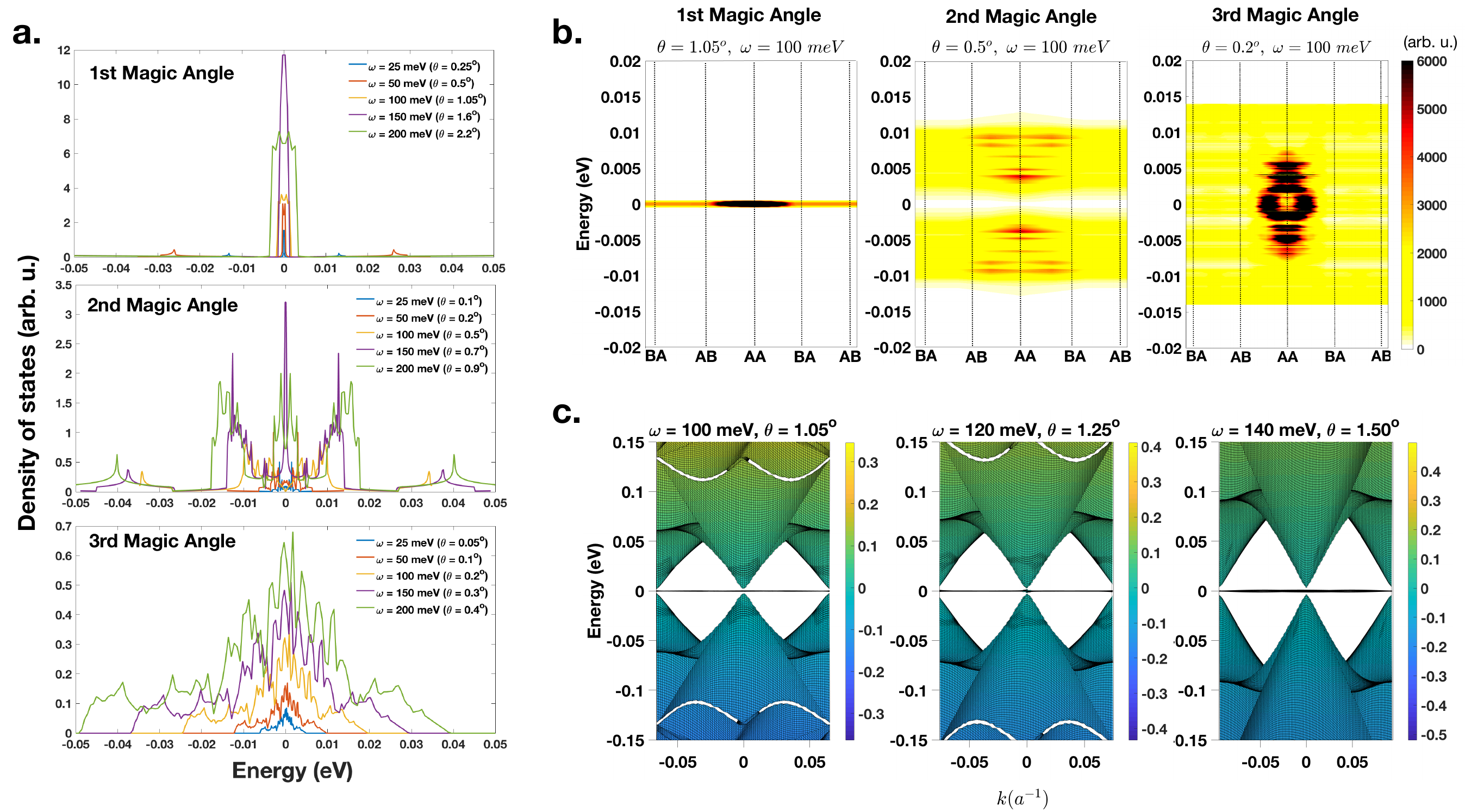} 
\end{center}
 \caption{
 (Color online) {\bf a.}
 Density of states (DOS) of the flatbands corresponding to the first (top panel), second (middle panel) and third (bottom panel) 
 magic angles for different values of interlayer tunneling $\omega$.
 We observe a progressive increase in the DOS maxima and width variations for larger twist angle flat bands due to the increase in the moir\'e Brillouin zone area.
 {\bf b.} Local density of states evaluated for the first, second and third magic angles along lines connecting different local stacking configurations.
 While the wave functions still localize at AA stacking regions the flatbands bandwidth widen for higher order magic angles. 
 {\bf c.} Band structure plots for different interlayer coupling strength and progressive increase in the magnitude of the first magic angles
 due to enhancement of interlayer tunneling achievable applying pressure. 
    }
 \label{figure3}
\end{figure*}
\bigskip
%-- Twist angle range --
The twist angles range examined in the phase diagram lie between 0.05$^{\circ}$ and 2.5$^{\circ}$. 
Convergence of the eigenvalues and eigenvectors of our continuum model can be expected when truncation of the moir\'e reciprocal lattice 
vector is of the order of $k   \sim   2 \omega / (\sqrt{3} a \left| t_0 \right|)$~\cite{bistritzer2011}, requiring a larger cutoff when $\omega$ is larger.
We have used a cutoff in momentum space for a radius of about six moir\'e reciprocal lattice vectors $6 G_1=24\pi \theta/ (\sqrt{3} a)$
using fixed Hamiltonian matrix sizes of 676$\times$676 to obtain the phase diagram in Fig.~\ref{figure1}, 
which should be valid for sufficiently large $\theta \gtrsim \omega / ( 12 \pi \left| t_0 \right| )$ 
or in the limit of small $\omega$.
Our model assumes rigid in-plane lattices although for systems with small twist angles $\theta \lesssim 0.5^{\circ}$ one 
expects structural instabilities associated with the commensuration moir\'e strains due to the reduction of the elastic 
energies that scale with the twist angle as $\propto \theta^2$~\cite{jung2015}.

%-- Pressure range ---
Pressure can be varied in a continuous manner even after device fabrication to modify the magnitude of interlayer tunneling. 
Recent experimental progress that make use of a hydraulic press allowed to achieve continuously variable pressures of up to 2.5~GPa for hBN encapsulated 
graphene devices while values as large as 5~GPa might be achievable by optimizing the design of the press,
and even larger pressure would be achievable with diamond anvil cells.
These values are still safely below pressures %of 9~GPa 
where structural phase transitions from graphite to diamond can start taking place~\cite{gpressure1,gpressure2}.
The relationship between the interlayer coupling parameter $\omega$ and pressure $P$ is obtained by combining 
Fourier transformed interlayer hopping data between maximally localized Wannier functions obtained from LDA ab initio calculations
evaluated at three fixed interlayer separation distances of $c = 3.35, \, 3.2, \, 3.1~\AA$ for different interlayer stacking.
The pressures for different interlayer distances at $AB$ and $BA$ stacking configurations within the LDA
are given by $P=0, \, 2.01$ and $3.45$~GPa, and larger values for AA stacking of
$P= 2.09, 4.80, 7.73$~GPa within LDA consistent with the calculations in Ref.~\cite{leconte2017},
which can be fitted with the Murnaghan equation of state $P(V) = (K_0/K'_0) \left[ (V/V_0)^{-K'_0} - 1 \right]$ ~\cite{murnaghan}
or the third order Birch-Murnaghan equation of state~\cite{birch} using the parameters listed in Table~\ref{table1}.
By assuming approximately equal distribution of AB, BA and AA stacking areas, and averaging the values of interlayer tunneling at the Dirac 
point for different stacking configurations we obtain a polynomial fit for the relationship between pressure and averaged interlayer tunneling $\omega$ through
%the relationship between the pressure and the averaged interlayer tunneling through a quadratic polynomial fit of the form
\begin{eqnarray}
P = A\, \omega^2 + B\,  \omega + C,
%$P = A\, (\omega - \omega_0)  $
\end{eqnarray}
%where $A=62.57$~GPa/eV and $\omega_0=0.098$~eV is the tunneling at $P=0$.
whose numerical parameters are $A=455.5$~GPa/(eV)$^2$, $B=-71.05$~GPa/eV, $C=3.281$~GPa and 
$\omega_0=0.098$~eV is the tunneling for $P=0$.
%We have neglected the relative shrinking of $AA$ local stacking regions expected for small twist angles.
Because the tunneling is weaker for larger interlayer distances our % tunneling value of  
$\omega_0$ % % \sim 0.098$~meV 
consistent with relaxed LDA results is smaller than $\omega = 0.113$~eV obtained from the average 
of rigid graphene bilayers separated at fixed $c=3.35~\AA$~\cite{jung2014}. 
For sake of definiteness we will use for our zero pressure continuum model Hamiltonian the parameters $t_0=-2.6$~eV and $\omega=0.1$~eV 
that leads to electronic structure results closely similar to those obtained in Ref.~\cite{bistritzer2011}.
\begin{table}[] 
    \begin{tabular}{|c  | c c c | c c c  | c c c |}
      \hline
       &  &  AA & &  & AB & &  & BA &  \\ 
         \hline 
        &  LDA & & RPA &  LDA & & RPA & LDA & & RPA  \\ 
         \hline
     $K_0$  \input{MurnaghanK0.tex} \\
     $K_0^\prime$ \input{MurnaghanK0p.tex} \\
     $B_0$ \input{MurnaghanB0.tex} \\
     $B_0^\prime$ \input{MurnaghanB0p.tex} \\
         \hline
    \end{tabular} 
 %   \captionsetup{justification=centerlast}
    \caption{Bulk moduli $K_0$ and $K_0^\prime$ for Murnaghan and $B_0$ and $B_0^\prime$ in kbar units for the 
    third-order Birch-Murnaghan equations of state consistent with the interlayer potentials obtaiend in Ref.~\cite{leconte2017}
    obtained from ab initio calculations for different local stacking configurations. 
      }
     \label{table1}
\end{table}
The weaker interlayer tunneling regime $\omega < 0.1$~eV in the phase diagram might be achievable in systems whose average interlayer distances can be reduced
either through intercalation of ions or addition of a barrier hBN. The use of an intercalation hBN spacer layer between the top and bottom graphene layers 
to reduce interlayer tunneling can prevent the 
structural instabilities for small twist angles that are present when both graphene layers are in direct contact.

\section{Density of states}
The sharp increase in the density of states (DOS) and its width at the flatbands relative to Coulomb interaction strength
determines how prone the system is towards the development
of broken symmetry phases. Here we analyze the impact that the  interlayer coupling strength $\omega$ has in the density 
of states (DOS) when its value is increased along the flatband line equation given in Eq.~(\ref{mangle}) 
for a few select twist angles and interlayer tunneling $\omega$. 
The increase of the interlayer tunneling parameter $\omega$ expands the energy range around which a given Bloch state 
in one layer can scatter to the other layer and affects the eigenvectors associated with the flatbands.
An immediate effect of having larger magic twist angles is that we can expect an enhancement of the flatband DOS 
due to the increase of mBZ area proportionally to $\theta^2$, making more states accessible in momentum space,
because more electrons per unit area %moir\'e supercell 
are required to fill the same number of the moir\'e flatbands.
The reduction in the moir\'e pattern length implies a relative enhancement of the effective Coulomb interaction scale 
$U_{\rm eff}= e^2 / (4 \pi \varepsilon_0 \varepsilon_r \ell_M ) \sim e^2 \theta / ( 4 \pi \varepsilon_0 \varepsilon_r a ) $
where $\ell_M \sim a/ \theta$, and therefore an increase from $\theta \sim 1^{\circ}$ to $1.5^{\circ}$ leads to already a 50\% 
increase. 
We show in Fig.~\ref{figure3} the density of states and the local density of states associated with the charge neutrality 
flatbands for different values of twist angle $\theta$ and interlayer tunneling $\omega$
where we can observe a steady increase in the flatband DOS either for its peak height and width for progressively 
larger twist angle sizes. 
The broadening of the flatbands for smaller magic angles widens of the energy range around 
which the LDOS concentration takes place, reflecting the increase of interlayer coherence. 

\section{Summary and discussions}
In summary we have investigated the phase diagram map of the low energy bandwidth evolution in twisted bilayer graphene as a function 
of parameters in the continuum model Hamiltonian including the Fermi velocity $\upsilon_{\rm F}$, 
the twist angle $\theta$ and the interlayer tunneling parameter $\omega$ in search of the phase space where we can achieve 
bandwidth minima, in particular when the interlayer tunneling is enhanced by means of external pressure. 
The flatbands for the continuum Hamiltonian can be summarized in a single line equation
relating the minimum bandwidth magic twist angle with interlayer tunneling, 
and is inversely proportional to the Fermi velocity or intralayer hopping.
Our calculations indicate that by applying pressures on the order of $\sim$2~GPa achieved in recent experiments
through a hydraulic press~\cite{yankowitz} it should be possible to access the first magic angle around $\theta \sim 1.5^{\circ}$ 
which is considerably greater than $\sim 1^{\circ}$ and therefore should 
have considerably greater structural stability that can be altered by the moir\'e strains
and have enhanced effective Coulomb interaction energy scales $U_{\rm eff} \propto \ell^{-1}_M \propto \theta$ thanks to 
the reduction in the moir\'e pattern size. 
Hence, we can envision that application of pressure in twisted bilayer graphene nanodevices 
to achieve larger magic angles should considerably facilitate access to flatbands and electron-electron interaction driven ordered phases.

\section{Acknowledgement}
Support from the Korean National Research Foundation is acknowledged for  B.L.C. through NRF-2017R1D1A1B03035932, 
for N. L. through NRF-2018R1C1B6004437 and for J.J. through NRF-2016R1A2B4010105.  
N.L. and J.J. also acknowledge support by the Korea Research Fellowship Program through the National Research 
Foundation of Korea (NRF) funded by the Ministry of Science and  ICT (KRF-2016H1D3A1023826).
A.H.M. acknowledges financial support by the Army Research Office (ARO) under contract W911NF-15-1-0561:P00001, and by the Welch Foundation under grant TBF1473.

\end{document}

%% file: MurnaghanK0.tex
&322& &353&308& &358&308& &358

%% file: MurnaghanK0p.tex
&10.88& &12.59&12.51& &12.02&12.51& &12.02

%% file: MurnaghanB0.tex
&239& &331&302& &354&302& &354

%% file: MurnaghanB0p.tex
&14.27& &15.00&14.25& &13.48&14.25& &13.48